\documentclass[9pt,twocolumn,twoside]{opticajnl}
\journal{opticajournal} 
\setboolean{shortarticle}{false}

\usepackage{graphicx}
\usepackage{dcolumn}
\usepackage{bm}
\usepackage{xcolor}
\usepackage{upgreek}
\usepackage{bm}
\usepackage{lineno}
\usepackage{mathtools}
\usepackage{siunitx}

\usepackage{hyperref}

\title{Observation of spatially structured Montgomery effect in free space}

\author[1,2,*]{Murat Yessenov}
\author[1,3] {Luca Sacchi}
\author[1] {Alfonso Palmieri}
\author[4] {Layton A. Hall}
\author[2] {Ayman F. Abouraddy}
\author[1,$\dag$] {Federico Capasso}

\affil[1]{Harvard John A. Paulson School of Engineering and Applied Sciences, Harvard University, Cambridge, MA, USA}

\affil[2]{CREOL, The College of Optics \& Photonics, University of Central Florida, Orlando, Florida 32816, USA}

\affil[3]{Department of Physics, ETH Zürich, Zürich, Switzerland}

\affil[4]{Materials Physics and Applications - Quantum Division, Los Alamos National Laboratory, Los Alamos, NM 87545, USA}

\affil[*]{yessenov@seas.harvard.edu}
\affil[$\dag$]{capasso@seas.harvard.edu}

\begin{abstract}
We report the first direct observation of the spatially structured Montgomery effect, a lensless self-imaging phenomenon that generalizes the Talbot effect to aperiodic structures, unfolding repeated tightly focused spots ($\sim10$\;\unit{\um}) in free space. Using a dynamic optical hologram to discretize radial spatial frequencies, we demonstrate self-imaging at distances ranging from 30 to 100 mm. Our method independently controls the focal spot size and self-imaging period, enabling dynamic three-dimensional light patterns. We also show the arbitrary tunability of the transverse profile by demonstrating revivals of Laguerre-Gaussian, Hermite-Gaussian, Ince-Gaussian modes, and Airy beams. These findings open opportunities for multi-plane microscopy, optical atom traps, and quantum atomic systems. 
\end{abstract}

\setboolean{displaycopyright}{false}

\begin{document}

\maketitle
\section{Introduction}
Lens-less self-imaging is an interference phenomenon whereby a coherently structured wave field is reproduced at discrete distances. Talbot effect -- the first observation of self-imaging via optical illumination of a grating \cite{Talbot36PM} -- is the product of a near-field diffraction phenomenon from \textit{periodic} slits endowed with axial revivals separated by the self-imaging distance $z_{\mathrm{s}}$ \cite{Rayleigh1881xxv,Winthrop1965JOSA}. Over the years, the Talbot effect has been utilized in lithography \cite{Solak2011OE}, metrology \cite{Du2024PRL}, image processing \cite{wen2013AOP}, and more recently, in atom trap arrays \cite{Huft2022PRA}. The temporal Talbot has been shown to manifest self-imaging of pulse trains in dispersive media \cite{jannson1981JOSA}, and space-time Talbot \cite{Hall202APL} and veiled Talbot effects \cite{Yessenov2020PRLVeiled} have been observed in free space via spatiotemporally structured light \cite{Shen2023Roadmap}. 
However, a \textit{periodically} distributed intensity pattern in the transverse plane associated with the Talbot effect limits their utilization in some applications, such as multiplane fluorescence imaging \cite{Beaulieu2020NatMet}, whereby off-center intensity peaks may lead to unwanted excitation. For applications in optical trap arrays \cite{Schlosser2023PRL}, the Talbot effect’s widely \textit{distributed} axial intensity pattern can lead to particle displacement outside the self-imaging planes and reduced tweezer power per trap. For applications requiring micron-scale focusing, self-imaging is constrained to a few axial planes due to the near-field nature of the Talbot effect \cite{Di2004JOSA}.

\begin{figure*}[t!]
    \centering
    \includegraphics[width=17.6cm]{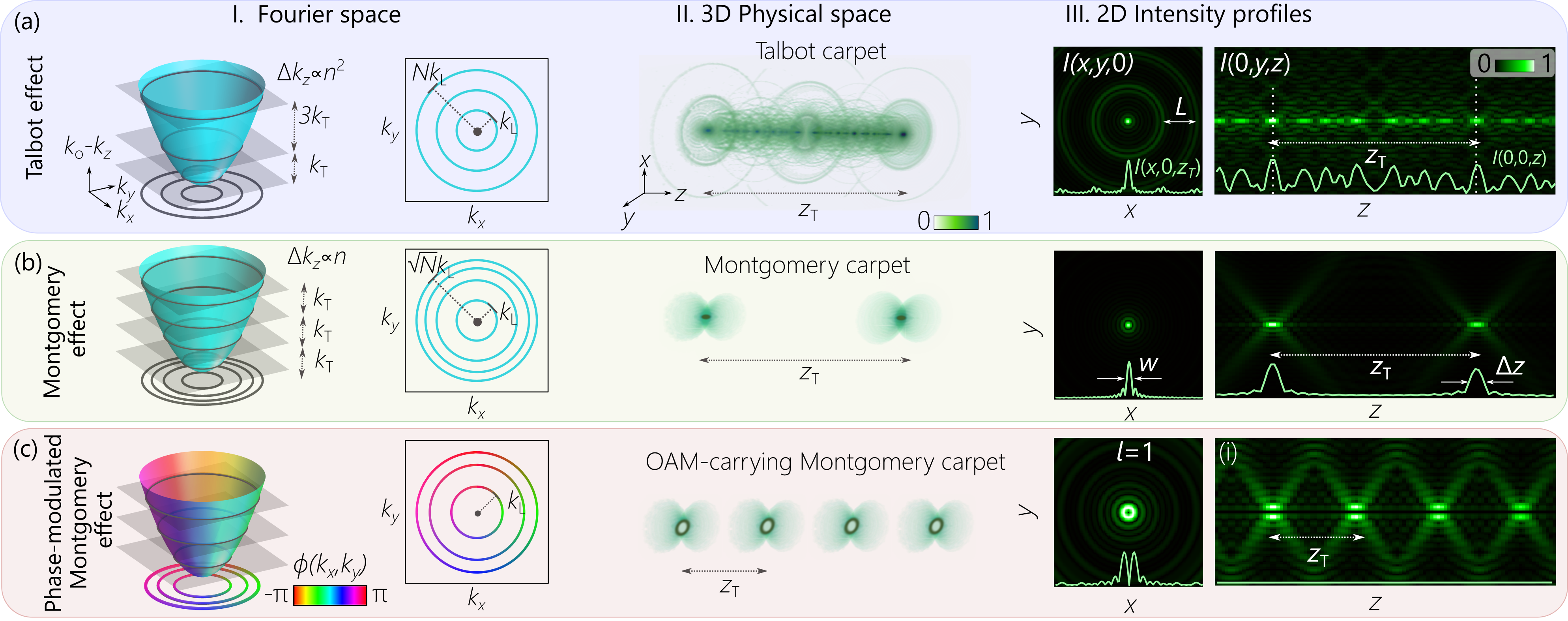}
    \caption[Talbot–Montgomery self-imaging]{\textbf{The concept of the Talbot and Montgomery effects}. Column I displays the Fourier space in which the spatial spectra of the corresponding field lay on the intersections of horizontal iso-$k_z$ planes and the paraboloid, which represents the dispersion relation $k_{o}\!-\!k_{z}\!\!=\!\!(\!k_x^2\!+\!k_y^2)/2k_{o}$;  the colormap of the surface and the lines represent the spectral phase $\phi(k_x,k_y)$. Column II depicts the intensity distributions $I(x,y,z)$ of the corresponding field in 3D physical space, and Column III cross-sections at $z\!=\!z_{\mathrm{s}}$ and $x\!=\!0$. 
    (a) The Talbot effect produced via a \textit{periodic} sampling of radial frequencies $k_{r}$.  The Montgomery effect generated from a \textit{aperiodic} sampling of radial frequencies $k_{r}$ with (b) a constant spectral phase and (c) spiral phase $\widetilde{\psi}(k_{r},\chi)\!=\!e^{i\chi}$. Here $k_{r}\!=\!\sqrt{k_x^2\!+\!k_y^2}$ and $\chi\!=\!\arctan{(k_y/k_x)}$. }
    \label{fig:concept}
\end{figure*}

A lesser-known yet potentially more versatile self-imaging phenomenon -- the Montgomery effect -- can produce axial revivals of \textit{aperiodic} structures, generalizing the concept of self-imaging beyond periodic patterns \cite{Montgomery67JOSA}. The fundamental distinction between these self-imaging effects lies in the spatial spectrum. The Talbot effect arises from linearly discretized transverse spatial frequency  $k_{\perp}\!\sim\!n$ ($n\!\in\mathbb{N}$) (Fig.~\ref{fig:concept} (a), Column I), whereas for the Montgomery effect, transverse spatial frequency is discretized with square-root sampling $k_{\perp}\!\!\sim\!\!\sqrt{n}$ (Fig.~\ref{fig:concept} (b), Column I) \cite{Montgomery67JOSA,lohmann_fractional_2005}. Due to the aperiodic sampling in the spatial spectrum, the Montgomery is clear from the periodically distributed transverse profile and can sustain self-imaging over longer propagation distances at a high numerical aperture (NA). Unlike the Talbot effect, which yields a fractal-like tapestry of light known as the Talbot carpet (Fig.~\ref{fig:concept} (a), Column II), the Montgomery effect produces periodic revivals of axially \textit{localized} fields without global structure (Fig.~\ref{fig:concept} (b), Column II). Moreover, localized intensity profiles at the imaging planes, overlaid with phase and amplitude modulation in Fourier space, enable arbitrary structuring of the spatial profile. For example, to introduce orbital angular momentum into the beam while maintaining its self-imaging property (Fig.~\ref{fig:concept} (c)). Such tight localization and control over the spatial profile make the Montgomery pattern more attractive for applications requiring discrete replication of well-defined beam profiles, such as optical manipulation of particles \cite{Dholakia2011NP}, two-photon microscopy \cite{helmchen2005NatMet}, and multi-photon microscopy \cite{Konig2000Book}. 
Previous experimental observations of the Montgomery effect primarily focused on imaging applications, demonstrating axial revivals of an image via periodic \cite{indebetouw_polychromatic_1988} and aperiodic sampling of spatial frequencies \cite{Indebetouw1992JOSAA}. These demonstrations relied on hard amplitude masks to discretize spatial frequencies, which lack dynamic controllability and are associated with a strong filtering effect. Subsequent efforts exploited computer-generated holograms to observe the Montgomery effect \cite{jahns_montgomery_2003}, but were limited to rudimentary designs that reported transverse profiles at a few planes. The 'Montgomery carpet' -- the evolution of its intensity profile over propagation  -- has not been directly observed to date. Additionally, control over its characteristic features, such as beam size, depth of focus, and the effects of spectral intensity and phase modulations, has not yet been demonstrated.

Here, we present the observation of an optical Montgomery carpet, manipulating the propagation structure of tightly focused light beams by dynamically tuning the spatial amplitude and phase structure. This is accomplished by utilizing a user-controlled holographic phase mask displayed on a spatial light modulator (SLM). Our technique enables independent and versatile tunability of the beam profile and self-imaging distance of the Montgomery carpet, demonstrating transverse localization down to $\sim\!\!10$\;\unit{\um} at self-imaging distances from 30~mm to 100~mm. By carefully selecting the amplitude and phase profile of the spatial spectrum, we demonstrate the self-imaging of different Hermite-Gaussian modes, Laguerre-Gaussian modes endowed with orbital angular momentum (OAM) \cite{allen1999PiO}, generalized Ince-Gaussian modes with controlled ellipticity \cite{Bandres2004OL}, and a one-dimensional (1D) Airy beam traveling in a curved trajectory \cite{Siviloglou07OL}. Indeed, this methodology can, in principle, be extended to any spatially structured light forms, opening new avenues for the exploration of self-imaging with structured light \cite{Yessenov22AOP, Hu2025NP}. We also generated the three-dimensional (3D) Talbot effect in cylindrical coordinates by discretizing the radial frequencies periodically, presenting a tapestry of light distributed in a 3D volume. We believe the Montgomery effect, endowed with a localized self-imaging pattern, may serve as a platform for more controllable and high-precision optical manipulation techniques for atomic arrays, particularly for nuclear spin qubits \cite{Kusano2025PhysRevRes}, three-dimensional single-atom qubits \cite{Schlosser2023PRL}, ring traps for quantum gases \cite{De2021JoP, Ryu13PRL}, and manipulation of Bose-Einstein condensates (BECs) \cite{Ramanathan11PRL}. Moreover, the Montgomery effect with suppressed side-lobes may potentially be a vital tool for simultaneous multi-plane imaging \cite{Beaulieu2020NatMet} and volumetric imaging of synaptic transmission \cite{Chen2024NatMet}. Beyond these potential applications, the localization and self-imaging phenomena associated with the Montgomery effect reveal a rich repertoire of behavior intrinsic to wave fields with carefully crafted spatial discretizations in Fourier space \cite{Hornberger2012RevModPhys, Zhang2014NC}.

\section{Self-imaging condition in free space}
 We first outline the theoretical basis for the Montgomery effect. Throughout, we consider scalar paraxial monochromatic fields of frequency $\omega_{o}$ (wavenumber $k_o=\omega_o/c$) propagating along the $z$-axis, where $c$ is the speed of light.  Writing the field in cylindrical coordinates $(r,\varphi,z)$ as $E(r,\varphi,z;t)=e^{i(k_{o}z-\omega_{o}t)}\,\psi(r,\varphi,z)$, the envelope is expressed by
\begin{equation}\label{Eq:Envelope}
\psi(r,\varphi,z)\!=\!\!\int_{0}^{\infty}\!\!\!\!\int_{0}^{2\pi}\!\!\!\!\!k_{r}\widetilde{\psi}(k_{r},\chi)\,e^{ik_{r}r\cos(\varphi\!-\!\chi)}e^{\!-i(k_{o}\!-\!k_{z})z}\,d\chi\,dk_{r},
\end{equation}
where $\widetilde{\psi}(k_r,\chi)$ is the spatial spectrum, $k_z$ and $k_{r}\!=\!\sqrt{k_x^2\!+\!k_y^2}$ are the axial and radial spatial frequencies, respectively, which are related by $k_{z}\!=\!k_{o}\!-\!k^{2}_{r}/2k_{o}$ in the paraxial regime \cite{SalehBook07}, $\chi\!=\!\arctan{(k_y/k_x)}$, and  $k_x,k_y$ are the transverse spatial frequencies. The phase term $\xi\!=\!(k_{o}\!-\!k_{z})z$ in Eq.~\ref{Eq:Envelope} dictates the evolution of the field profile over propagation along $z$, and depending on the spatial spectrum $\widetilde{\psi}(k_{r},\chi)$, the diffraction pattern varies significantly. For example, Bessel beams \cite{Durnin87PRL} and cosine waves \cite{Lin2012PRL}, whose spatial spectra are comprised of $k_{z}\!=\!\mathrm{const.}$, propagate diffraction-free in free space \cite{FigueroaBook14}, which is the monochromatic case of a more general class of broadband propagation-invariant space-time wave packets \cite{Yessenov19PRA,Yessenov22AOP}. On the other hand, self-imaging -- periodic revival of the field structure $\psi(r,\varphi,0)$ at discrete axial planes $z\!=mz_{\mathrm{s}}$ ($m\in\mathbb{Z}$) -- is attained for plane wave components $k_z$ that satisfy \cite{Supplementary}:
\begin{equation}
    \xi\!=\!(k_{o}\!-\!k_{z})z_{\mathrm{s}}\!=\!2\pi m. 
\end{equation}

For the Talbot effect (Fig.~\ref{fig:concept} (a)), the spatial spectrum of the field is composed of plane-wave components that can be represented by \cite{Winthrop1965JOSA,Supplementary}
\begin{equation}
    \mathrm{\textit{Talbot}}:\qquad k_{r}(n)=k_{\mathrm{L}}n; \qquad k_{z}(n) =k_{o}-k_{\mathrm{s}} n^2; \label{Eq:kz_krTalbot}
\end{equation}
where $k_{\mathrm{s}}\!=\!2\pi/z_{\mathrm{s}}$ and $k_{\mathrm{L}}\!=\!\sqrt{2k_{o}k_{\mathrm{s}}}\!$ are the sampling coefficients in $k_z$ and $k_{r}$ respectively, and $n\in\mathbb{N}$ (Fig.~\ref{fig:concept} (a), Column I); see Supplementary Material Section I.A \cite{Supplementary}. The Montgomery self-imaging (Fig.~\ref{fig:concept}(b)) is composed of discrete plane-wave components \cite{Montgomery67JOSA,Supplementary}
\begin{equation}
    \mathrm{\textit{Montgomery}}: k_{r}(n)=k_{\mathrm{L}}\sqrt{n};\; k_{z}(n) =k_{o}-k_{\mathrm{s}} n. \label{Eq:kz_krMontg}
\end{equation}
The corresponding spatial spectrum of the field composed of such discrete spatial frequencies may be represented by
\begin{equation}\label{Eq:Spectrum_Mont}
    \widetilde{\psi}(k_r,\chi) = \widetilde{A}(k_{r},\chi) \sum_{n=1}^{N} \delta(k_r -k_{\mathrm{L}}\sqrt{n}),
\end{equation}
where $\delta(\cdot)$ is the Dirac delta function, $\widetilde{A}(k_{r},\chi)$ is a complex overlay function whose physical importance will be explained in the next paragraph, and $N$ is the maximum number of rings in Fourier space (Fig.~\ref{fig:concept} (b), Column I). The fundamental distinction between the Montgomery and Talbot effects in spectral space is the discretization functions of $k_{r}(n)$ and $k_{z}(n)$ in Eq.~\ref{Eq:kz_krTalbot} and Eq.~\ref{Eq:kz_krMontg} (Fig.~\ref{fig:concept}, column I). The spatial spectrum of the field for the Talbot effect is composed of concentric rings of linearly increasing radius $k_{r}(n)$ separated by $k_{\mathrm{L}}$, and the corresponding sampling in $k_{z}$ is quadratic, as depicted in Fig.~\ref{fig:concept} (a), column I. In the case of the Montgomery effect, the linear sampling is rather in axial frequency $k_{z}$ (with periodicity $k_{\mathrm{s}}$). Consequently, the spatial spectrum is composed of concentric rings with square-root sampling in $k_{r}(n)$,  Montgomery rings,  where the radius of the first ring is $k_{\mathrm{L}} $(Fig.~\ref{fig:concept} (b), column I).  

Such subtle modifications in the spatial spectrum of the field lead to stark consequences in physical space (Fig.~\ref{fig:concept}, columns II-III). First, the intensity distribution $I(x,y,z)\!=\!|\psi(x,y,z)|^2$  of the Montgomery effect portrays tight transverse and axial localizations, whereas the intensity in the Talbot carpet is widely distributed throughout the whole volume (Fig.~\ref{fig:concept} (a,b) column II). Particularly, in the Talbot effect, a periodic structure emerges (with periodicity $L\!=\!2\pi/k_{\mathrm{L}}$) at the Talbot plane, which is not observed in the Montgomery effect due to aperiodic sampling in $k_{r}$ (Fig.~\ref{fig:concept} (a,b) column III). Notably, the self-imaging distance for both effects remains the same, as given by the expression for the self-imaging distance $z_{s}\!=\!2L^2/\lambda_{o}$. However,  $L$ cannot be directly defined for the Montgomery effect; therefore, we can rewrite the expression as $z_s\!=\!\frac{8\pi^2}{\lambda k^{2}_{\mathrm{L}}}$. Second, by virtue of tight localization, the beam size $w$ and the depth of focus (DOF) $\Delta z$ can be defined for the Montgomery effect, where $w\!\propto\!\sqrt{\lambda_{o}z_{s}/N}$ and $\Delta z\propto z_{s}/N$ (Supplementary Material Section I.C \cite{Supplementary}). Both beam size $w$ and DOF $\Delta z$ can be controlled by tuning the number of rings $N$, while fixing the self-imaging distance $z_{\mathrm{s}}$ \cite{Supplementary}. Third, by controlling the overlay function $\widetilde{A}(k_{r},\chi)$, the beam profile and the corresponding Montgomery carpet can be arbitrarily tuned to portray various propagation dynamics while maintaining self-imaging. We illustrate this in Fig.~\ref{fig:concept} (c) by impinging a spiral phase $A(k_{r},\chi)\!=\!e^{i\ell\chi}$, where $\ell$ is the topological charge. In physical space, this generates a Montgomery carpet with overlaid OAM in the field (Fig.~\ref{fig:concept} (c), column II and III) and a topological charge conserved over propagation (Supplementary Material Section I.D \cite{Supplementary}). 
\begin{figure}[t!]
    \centering
    \includegraphics[width=8.7cm]{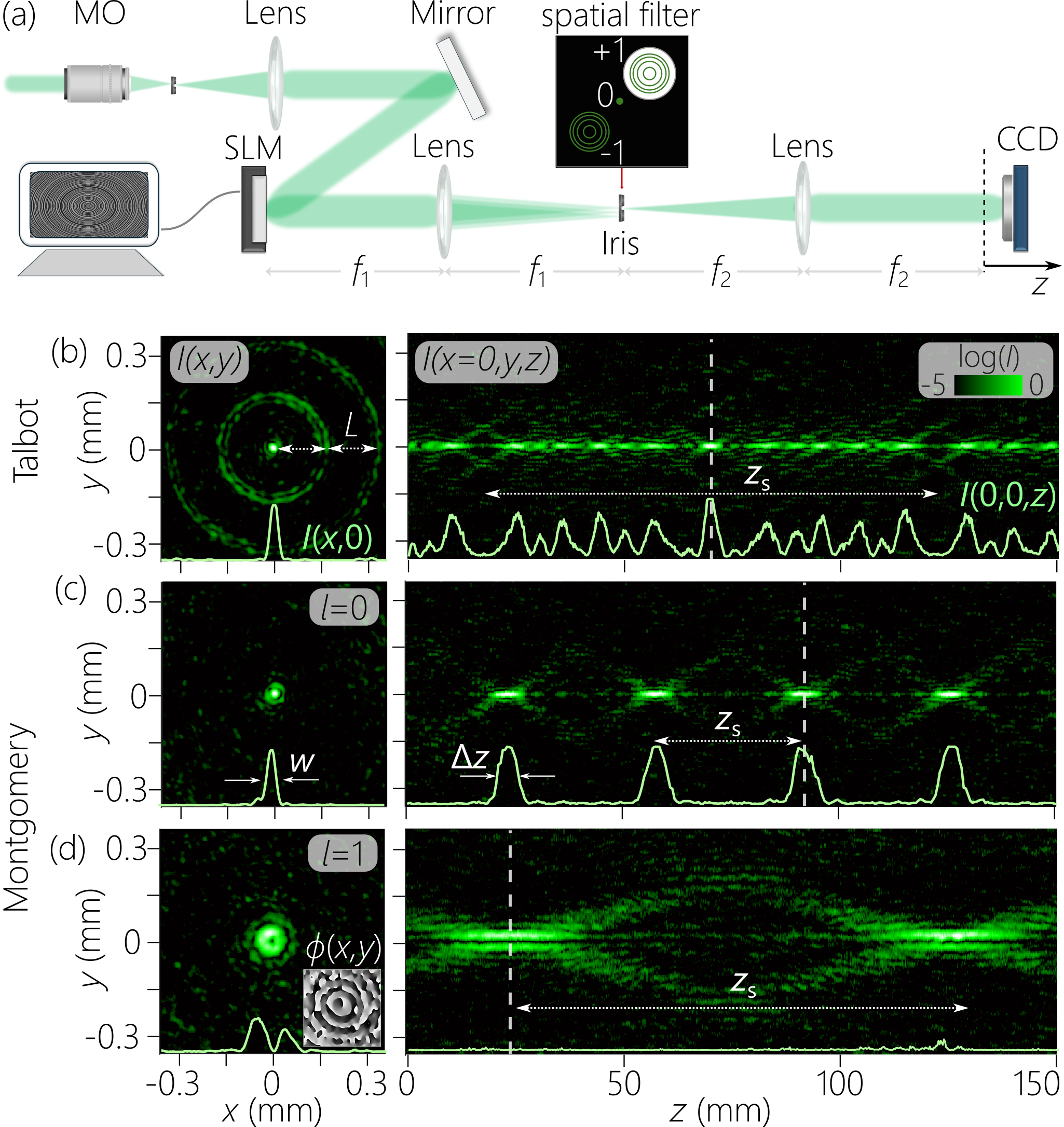}
    \caption{\textbf{Experimental realization of Talbot and Montgomery effects.} (a) Schematic of the synthesis and characterization setup consisting of a (i) phase‑only SLM to imprint the hologram, (ii) a 4-\textit{f} system with two spherical lenses ($f_{1}\!=\!f_{2}\!=\!250$ mm) with an (iii) iris at the Fourier plane for spatial filtering and to image the SLM plane, where (iv) a CCD camera axially scans the evolution of the intensity profile $I(x,y,z)$.  (b) Observation of the Talbot effect, Montgomery effect with (c) a flat phase profile and (d) a spiral phase profile $e^{i\ell\chi}$ ($\ell\!=\!1$). In (b-d)  left panels depict the transverse intensity profiles $I(x,y,z_{\mathrm{s}})$ at the Talbot planes (depicted by a dashed white line on the right panel), the right panel corresponds to the axial intensity profiles $I(y,z)$ at $x\!=\!0$. The inset on the left panel of (d) depicts the measured phase profile $\phi(x,y)$ at $z=25$~mm. We use a log-scale colormap for the intensity plots throughout the figure for clearer visualization.}
    \label{fig:setup}
\end{figure}

\section{Experimental setup}

\begin{figure*}[t!]
    \centering
    \includegraphics[width=17.6cm]{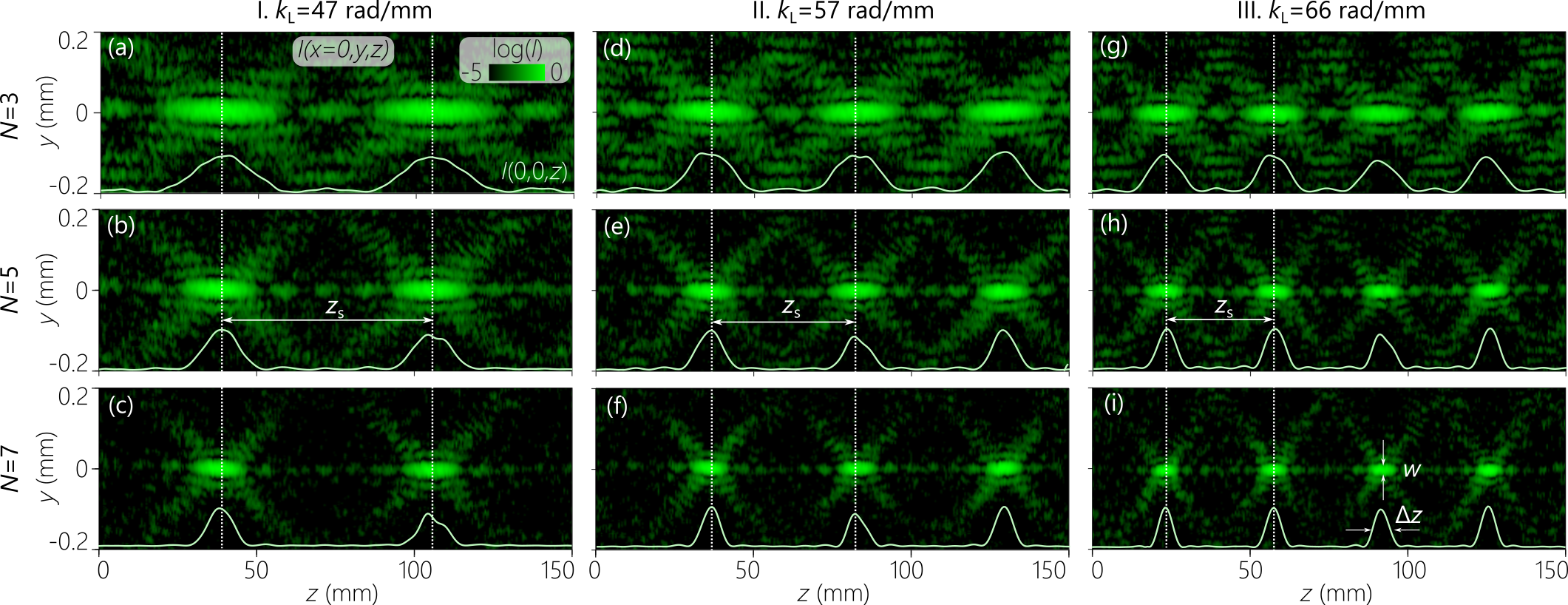}
    \caption{\textbf{Measured axial intensity distributions of Montgomery carpets.} Columns correspond to the radius of the Montgomery ring $k_{\mathrm{L}}\!=\!47$, 57, and 66\,rad/mm; rows correspond to the number of rings in the spatial spectrum $N\!=\!3$,  5, and 7. White line plots at the bottom of each panel represent normalized axial intensity profiles at the beam center $I(0,0,z)$. The beam size $w$ and depth of focus (DOF) $\Delta z$ are defined as the full width at half maximum of the intensity in the transverse and axial directions, respectively, as shown in (i). We use a log-scale colormap throughout the figure for clearer visualization.}
    \label{fig:Axial_scan}
\end{figure*}

The experimental setup for the synthesis and observation of the Talbot and Montgomery effects is sketched in Fig.~\ref{fig:setup} (a).  A continuous-wave laser beam at $\lambda=532\;\mathrm{nm}$ is expanded to \(15\;\mathrm{mm}\) in size and directed onto a reflective phase-only SLM under an oblique incidence of \(5^{\circ}\), where the target field profile $\psi(x,y,z\!=\!0)$ is displayed. Below, we explain in more detail how to generate the required SLM phase pattern. We then relay the SLM plane via a 4-\textit{f} system consisting of two spherical lenses ($f_{1}\!=\!f_{2}\!=\!250\,$mm) to the CCD camera to observe the Montgomery effect. At the Fourier plane of the 4-\textit{f} system, an adjustable iris is placed to filter out the zeroth and higher diffraction orders, passing the first diffraction order that carries the desired field profile (Supplementary Material Section II \cite{Supplementary}). 

To generate the SLM phase pattern $\Phi(x,y)$, we first formulate the spatial spectrum $\widetilde{\psi}(k_r,\chi)$ in Eq.~\ref{Eq:Spectrum_Mont} with the desired parameters $k_{L}$, the number of rings $N$, and the overlay function $\widetilde{A}(k_r,\chi)$, which is then converted to the field profile in physical space $\psi(x,y,z\!=\!0)$ via Eq.~\ref{Eq:Envelope}. By employing the complex-amplitude encoding strategy \cite{arrizon2007pixelated}, we convert the target field profile $\psi(x,y,z\!=\!0)$ into a computer-generated phase profile that is displayed on the SLM (see Supplementary Material Section II.A, B \cite{Supplementary}). 

To observe the Montgomery carpet, a CCD camera mounted on a motorized stage scans transverse intensity profiles $I(x,y)$ over the propagation distance of $z=150$~mm with a step of $0.25$ mm around the focal plane of the second lens. In Fig.~\ref{fig:setup}(b-d),  Fig.~\ref{fig:Axial_scan} and Fig.~\ref{fig:Fig5} we plot two-dimensional (2D) slices of the three-dimensional (3D) intensity pattern $I(x,y,z)$. We also employ the digital off-axis holography (ODH) technique \cite{Cuche2000AO} to extract the phase profile of an OAM-carrying beam at the first self-imaging plane shown in Fig.~\ref{fig:setup} (d), inset. Namely, the beam in Fig.~\ref{fig:setup} (d) and the reference plane-wave taken in front of the SLM are interfered at an oblique angle ($\sim2^{\circ}$) at the first imaging plane $z\approx25$~mm, where the image is captured with the CCD camera. Following the ODH algorithm, we perform a digital fast Fourier transform (FFT) to separate the constant background from the interference term. By digitally isolating the first diffraction order of the Fourier-transformed image, we access the interference term that contains the complex field. Finally, the inverse FFT of the centered first term gives the amplitude $|\psi(x,y)|$ and phase $\phi\{\psi(x,y)\}$ at that axial location $z$ (see \cite{Cuche2000AO,Yessenov22NC}  for more details). We also capture the spatial spectral intensity $I(k_x,k_y)\!=\!|\widetilde{\psi}(k_x,k_y)|^2$ (Eq.~\ref{Eq:Spectrum_Mont}) in the Fourier plane of the $4$-f system, which we plot in Fig.~S7 of the Supplementary Material \cite{Supplementary}. 

\section{Results}

\begin{figure}[b!]
    \centering
    \includegraphics[width=8.7cm]{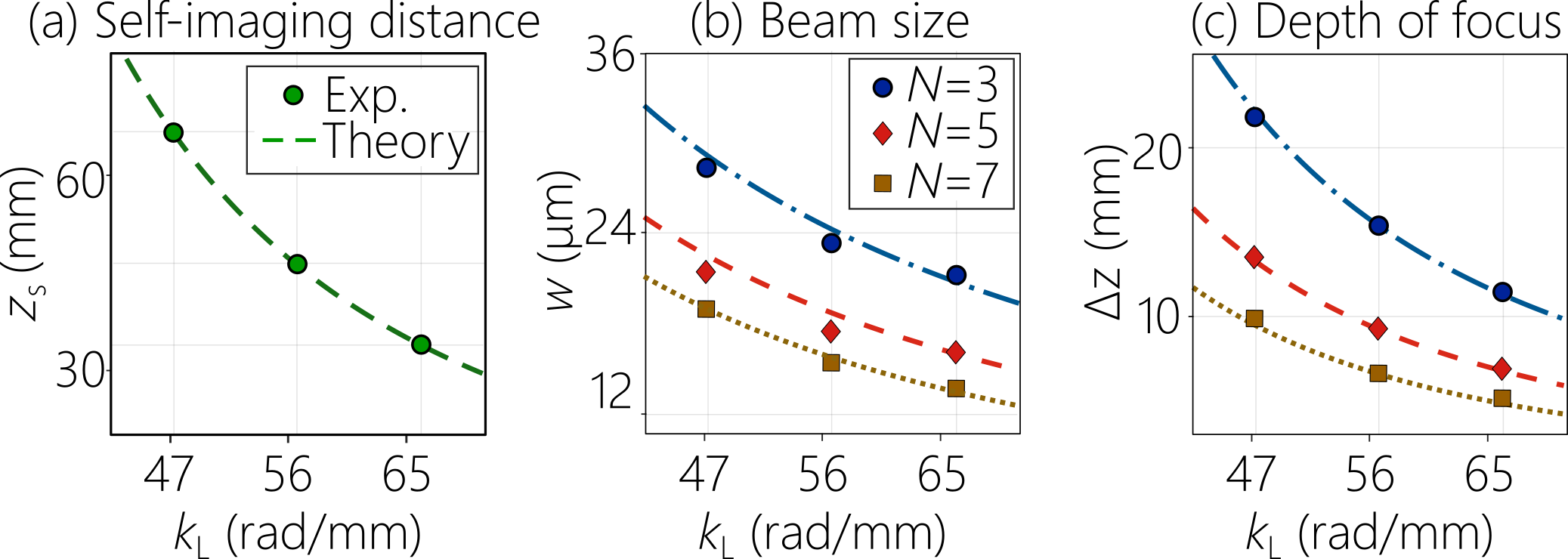}
    \caption{\textbf{Measured characteristic parameters of the Montgomery effect.} (a) self-imaging distance $z_{\mathrm{s}}$, (b) beam size $w$, and (c) the depth of focus (DOF) $\Delta z$
  as a function of $k_{\mathrm{L}}$. Points correspond to the measured data, and a line with the corresponding color indicates the theoretical plot.  In (b,c) different colors correspond to different numbers of rings $N$ in the Montgomery effect. }
    \label{fig:Fig4}
\end{figure}

\begin{figure*}[t!]
    \centering
    \includegraphics[width=17.6cm]{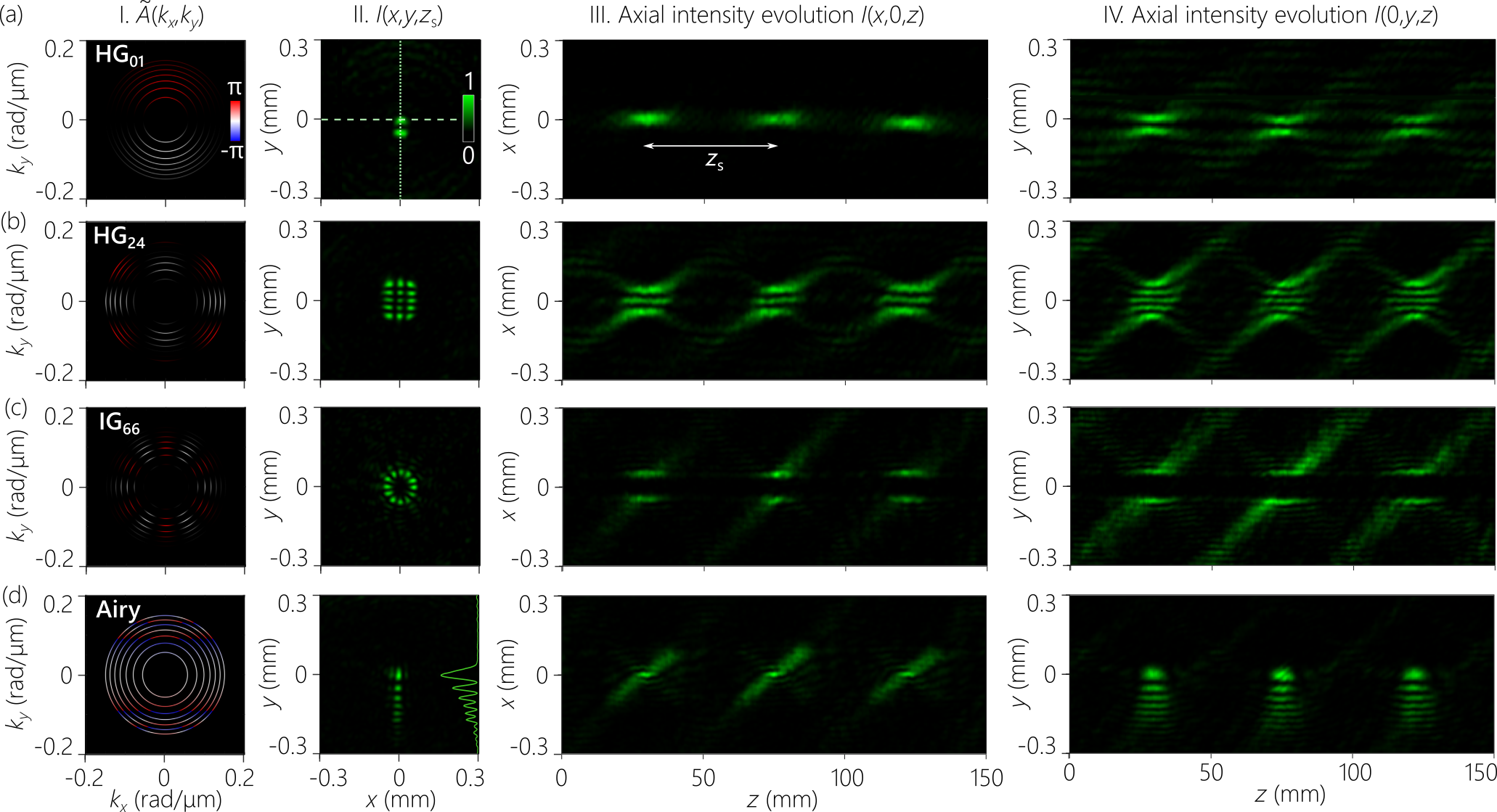}
    \caption{\textbf{Demonstration of the spatial structuring of the transverse profile of the Montgomery carpet}. Montgomery carpet with the transverse profile in the form of Hermite-Gaussian modes (a) HG$_{01}$ mode and (b) HG$_{24}$ mode; (c) Ince-Gaussian mode IG$_{66}$ mode with ellipticity $\epsilon=0.5$; (d) 1D Airy beam with cubic phase structure along the $y$-axis. Column I represents the calculated complex overlay function $\widetilde{A}(k_x,k_y)$, where the colormap depicts the phase. Column II corresponds to the transverse intensity profile $I(x,y)$ at the first self-imaging plane $z\approx25$~mm. Columns III and IV show the measured 2D intensity plots $I(x,z)$ at $y=0$ (horizontal dashed line in (a) Column II) and $I(y,z)$ at $x=0$ (vertical dotted line in (a) Column II), respectively.}
    \label{fig:Fig5}
\end{figure*}

\subsection{Observation of the Talbot and Montgomery carpets}
In Fig.\ \ref{fig:setup} (b), we show the experimental results of the Talbot effect. The transverse intensity profile $I(x,y)$ at the Talbot plane $z\!\approx\!75\,$mm (left panel in Fig.~\ref{fig:setup} (b)) reveals a central peak with a diameter of $w\!\approx\!16\,$\unit{\um}, accompanied by off-center rings separated by $L\!\approx\!168$\,\unit{\um}. The axial evolution $I(0,y,z)$ (right panel in Fig.~\ref{fig:setup} (b)) portrays the familiar 2D Talbot carpet with self-imaging at the planes separated by the self-imaging distance of $z_{\mathrm{s}}\!\approx\!105\,$mm. Notably, here we observed a radially-symmetric Talbot carpet in a 3D volume generated from concentric rings, which contrasts with conventional 2D Talbot carpets obtained from 1D or 2D gratings \cite{wen2013AOP} and the more recent observation of the angle–orbital angular momentum Talbot effect \cite{Hu2025NP}. Next, we carry out the measurements for the Montgomery effect with a similar beam size and self-imaging length of $z_{s}\!=\!34\,$mm, as shown in  Fig.~\ref{fig:setup} (c). The intensity is tightly localized in the transverse plane (left panel) as well as axially, with a beam diameter $w\!\approx\!18\;$\unit{\um} and a depth of focus (DOF) of $\Delta z\!\approx\!6$~mm (right panel). In Fig.~\ref{fig:setup} (d), we impose a phase profile $\widetilde{A}(k_r,\chi)=e^{i\chi}$, which imprints OAM with $\ell = 1$ in the Montgomery carpet. Such a phase profile manifests itself as a well-recognized donut-shaped intensity profile and a spiral phase profile in the transverse plane (Fig.~\ref{fig:setup}(d), left panel). Crucially, the $y$–$z$ intensity map (Fig.~\ref{fig:setup} (d), right panel) retains the tight axial confinement, demonstrating that the Montgomery self‑imaging is robust to additional wavefront structuring. We will further investigate the effects of the amplitude and phase overlay functions in the next section. 


In Fig.~\ref{fig:Axial_scan}, we experimentally demonstrate the versatility and independent tunability intrinsic to Montgomery self-imaging by exploring its axial evolution across different spectral configurations. We vary the radius of the first ring in the Fourier space $k_{\mathrm{L}}$ from $k_{\mathrm{L}}\!=\!47$\,rad/mm (Column I),  57\,rad/mm (Column II) to  66\,rad/mm (Column III), which is expected to result in self-imaging distances of $z_s=67.3$~mm, 45.7~mm, and 34.1~mm, respectively. The recorded intensity profiles reveal distinct self-imaging revivals, clearly confirming the robust axial periodicity of the Montgomery carpet across multiple planes. We then increase the number of rings $N$ from 3 to 7 (Rows I-III) for each $k_{\mathrm{L}}$. Concurrently, Fig.~\ref{fig:Axial_scan} highlights how incrementally increasing the number of spectral rings $N$ sharpens both transverse and axial localization. 

For quantitative comparison between the measured and theoretically expected characteristic features of the Montgomery carpet, in Fig~\ref{fig:Fig4} we plot the self-imaging distance $z_{s}$,  beam size $w$, and depth of focus (DOF) $\Delta z$ from the data in Fig~\ref{fig:Axial_scan}. The self-imaging distance $z_{\mathrm{s}}$ is governed solely by the first radial spectral frequency $k_{\mathrm{L}}$, as confirmed by our experimental data shown in Fig.~\ref{fig:Fig4}(a). The transverse beam size ($w$) and the depth-of-focus ($\Delta z$) can be precisely tuned by altering the number of spectral rings ($N$). This behavior is demonstrated clearly in Fig.~\ref{fig:Fig4}(b) and (c), respectively. Throughout Fig.~\ref{fig:Fig4} experimental measurements fit well with the theoretical predictions (see Supplementary material Section III \cite{Supplementary}).

\subsection{Spatial structuring of the transverse profile of the Montgomery carpet}

Up to this point, we have not tuned both the amplitude and phase structures of the Montgomery effect simultaneously. However, the existence of the overlaying function $\widetilde{A}(k_r,\chi)$ implies that we may implement structured light to the Montgomery effect, such that various forms of structured light may self-image.  To examine this, return to Eq. \ref{Eq:Spectrum_Mont} to explore more exotic self-imaging carpets that are available for the Montgomery effect. Considering the overlay function $\widetilde{A}(k_r,\chi)$, we may determine the carpet for any given overlay as follows:

\begin{equation}
    \psi(r,\varphi,z) = A(r,\varphi) \ast \psi_{M}(r,\varphi,z)
\end{equation}

\noindent where $A(r,\varphi)$ is the Fourier transform of the overlay function $\widetilde{A}(k_r,\chi)$, $\psi_{M}(r,\varphi,z)$ is the field for the Montgomery carpet, and $\ast$ denotes a convolution in both the $x$ and $y$ dimensions. This implies that we may convolute any spatial structure with the Montgomery carpet.

Therefore, to demonstrate this consequence, we will first consider one of the fundamental modes of structured light, the Hermite-Gaussian (HG) mode. Fig. \ref{fig:Fig5}(a) depicts the HG$_{01}$ with a Montgomery carpet having the parameters of $k_L = 57$~rad/mm and $N = 7$ rings. As a result, this produces a self-imaging distance of $z_{s} \approx 45$~mm, which is held constant across all configurations considered in Fig. \ref{fig:Fig5}. In the spatial frequency spectrum, $\widetilde{\psi}$ is calculated in column I, and we find that the Montgomery rings are overlaid with a $\pi$ phase shift between the top and bottom of the rings. As a result, in the self-imaging plane demonstrated in column II, the HG$_{01}$ mode is measured and appears as expected. The axial measurements are reported in columns III and IV, resulting in the self-imaging of the HG$_{01}$ over three planes. To further expand this result, we consider a higher order HG$_{24}$ in Fig. \ref{fig:Fig5}(b). Likewise, the same measurements are made across (b), resulting in the self-imaging of the initial formation and reconstruction of the HG$_{24}$.
To generalize this further, we consider the Ince-Gaussian (IG) mode. These modes are considered a hybrid between the HG and Laguerre-Gaussian (LG) modes \cite{Bandres2004OL}, which have found applications in a variety of areas \cite{Kotlyar25Book}, including particle trapping and manipulation \cite{Woerdemann2011APL}. In Fig. \ref{fig:Fig5}(c), we implement the IG$_{66}$ mode with an ellipticity of $\epsilon = 0.5$. As expected, Fig. \ref{fig:Fig5}(c) plots the resulting calculated spatial spectrum in I and the measured profile $I(x,y)$ in column II. Likewise, the reconstructions are observed over the propagation distance in columns III and IV.

Finally, to consider an exotic structure, we imprint the Airy phase across the Montgomery rings. The Airy beam is an infamous structure in optics due to its propagation invariant behavior and curved trajectory \cite{Siviloglou2007PRL}. Here, we utilize the Airy phase in Fig. \ref{fig:Fig5}(d) with a cubic phase of $\widetilde{A}(k_r,\chi)=\exp{\{i\left(\frac{k_r \cos{\chi}}{\kappa}\right)^3\}}$, where the spectral coefficient is $\kappa=1.1\,k_{\mathrm{L}}$, resulting in the Montgomery rings being overlaid with an Airy phase, as seen in column I. As a result, a 1D Airy pattern is formed in column II with a first lobe size of $\approx30~\unit{\um}$. In the axial propagation, the Airy pattern is self-imaged in columns III and IV. However, in IV, the characteristic bend of the Airy beam can be seen between the self-imaging planes. While these examples are not exhaustive, they demonstrate the unique ability of the Montgomery effect to produce singular reconstructions of arbitrary forms of structured light.

\section{Discussion}
The development of optical self-imaging itself enjoyed several historical revivals separated by decades of inactivity: from Talbot’s first observation \cite{Talbot36PM} to Lord Rayleigh’s theoretical formulation \cite{Rayleigh1881xxv}, to the Fresnel-image theory of Winthrop and Worthington \cite{Winthrop1965JOSA}, which conceptualized the modern-day understanding of the Talbot effect and paved the way for a plethora of applications that followed \cite{wen2013AOP}. Similarly, the Montgomery effect was experimentally realized \cite{indebetouw_polychromatic_1988,Indebetouw1992JOSAA,jahns_montgomery_2003} decades after its theoretical proposal \cite{Montgomery67JOSA}. However, these observations were narrowly focused on imaging applications \cite{indebetouw_polychromatic_1988,Indebetouw1992JOSAA} and have not comprehensively investigated the propagation properties and characteristic features of the Montgomery effect \cite{jahns_montgomery_2003}, precluding its application across various fields. Direct observation of the 'Montgomery carpet' -- dynamic evolution of the aperiodically sampled optical field -- presented in this work, therefore, may unveil fascinating properties of this effect. Crucially, quantitative analysis of the link between intrinsic parameters (sampling radius, number of rings) and extrinsic features (beam size, depth of focus, self-imaging distance) of the Montgomery carpet, as well as the experimental demonstration of the independent tunability of these parameters, provides an intuitive understanding and parameter selection for future applications. In combination with the versatile methodologies for spatially structuring light fields \cite{Forbes2021NP}, the Montgomery self-imaging effect offers a programmable self-imaging platform.  

Suppressed side lobes in the imaging plane of the Montgomery carpet, compared to the conventional Bessel beam, provide improved spatial resolution in fluorescence imaging, as recently reported in ref. \cite{Chen2024NatMet}. The Montgomery effect presented here offers more control over the depth of focus and beam size by tuning the number of Montgomery rings $N$, which may potentially further improve the resolution for the imaging of synaptic transmissions \cite{Chen2024NatMet}. Moreover, the axial repetition of highly focused light beams with dynamically tunable self-imaging distances $z_s$ may find applications in simultaneous multi-plane two-photon microscopy \cite{Beaulieu2020NatMet} and multi-Z confocal microscopy \cite{Badon2019Optica}. Within this work, we limited our studies to the paraxial regime, where square-root sampling of the spatial frequencies $k_r\!=k_{\mathrm{L}}\sqrt{n}$ is sufficient for self-imaging \cite{Montgomery67JOSA}. However, for the application in high-resolution microscopy, spot sizes of $\sim1~\unit{\um}$ are typically necessary, which is beyond the paraxial limit. Future work will focus on generalizing the self-imaging condition for the non-paraxial regime and investigating the most suitable synthesis methodologies for high-NA systems. 

Recent interest in single-atom trap arrays utilizing Talbot tweezer lattices underlines the versatility of self-imaging optical fields in quantum computing \cite{Schlosser2023PRL, Pause2024Optica}. By extending the 1D array presented here to 2D and 3D trap arrays, the 'Montgomery tweezers' with more localized power distribution may potentially produce a large number of single-atom traps with lower power requirements. Moreover, by reducing the thickness of each Montgomery ring in the Fourier space to a few microns, one could extend the Montgomery carpet to $\sim100$ self-imaging planes \cite{indebetouw_nondiffracting_1989}. Therefore, in principle, by carefully designing an array of $100\times100$ Montgomery carpets in the transverse plane, each extending to 100 imaging planes axially, it would be possible to produce 1 million trap arrays, vastly extending the capabilities of large-scale atom array systems \cite{chiu2025Nature}. We believe a natural candidate for the single-element implementation of such 3D Montgomery tweezers is meta-optics \cite{Khorasaninejad16Science, Dorrah2022Science}, whose recent technological advancements allow for the fabrication of highly efficient centimeter-scale meta-optics with subwavelength feature sizes \cite{Park2019ACS}. The optical tweezer architecture for the programmable atom trap arrays could include various combinations of meta-surfaces, forming the Montgomery rings, together with spatial light modulators for dynamic control of each 1D array via tuning the overlay function $\widetilde{A}(k_r,\chi)$, and acousto-optic deflectors for fast transportation \cite{bluvstein2024nature,chiu2025Nature}. 

Although we focused on scalar fields in this work, the concept could further be extended to achieve self-imaging of vector fields and polarization tunability across self-imaging planes using meta-optics \cite{Dorrah2021NC}. Polarization control of the Montgomery carpet via meta-optics could be particularly useful for applications in quantum telecommunications, where high polarization purity is imperative for efficient coupling in some neutral atoms \cite{chaneliere2006PRL}. Moreover, this toolbox offers an alternative route to realize Josephson vortices in Bose-Einstein condensates (BECs) \cite{Bazhan22PRL,RoditchevNP15,CaputoNP19} and to probe new dynamical regimes in BECs, atomtronics, and structured quantum systems \cite{AmicoAVS21}.

Combining the results presented here with the recent advances in spatially structured \cite{Forbes2021NP} and spatiotemporally structured light fields \cite{Shen2023Roadmap,Yessenov22AOP} opens up new avenues for the investigation of structured self-imaging effects in space, time, and jointly in space and time. 
Because the Montgomery effect and cylindrically symmetric structured fields both share the Bessel basis \cite{Dorrah2021NP}, previously demonstrated axial-encoding tools can be applied directly to the Montgomery effects, including the rotatum of light \cite{Dorrah2025SciAdv} and on-demand OAM transformation \cite{Dorrah2021NP}. Only very recently, novel space-time self-imaging phenomena have been observed in free space - veiled Talbot \cite{Yessenov2020PRLVeiled}, temporal Talbot effect \cite{Hall2021OL}, and space-time Talbot effect  \cite{hall2021APL}, some of which relate to the traditional Talbot effect via relativistic transformation \cite{Yessenov2023PRL}.

\section{Conclusion}
 We have demonstrated the spatially structured Montgomery self-imaging effect, encompassing a sequence of tightly localized and arbitrarily shaped optical beams over extended distances via dynamic holographic phase masks.  We have shown how intrinsic spectral parameters of the Montgomery ring map onto extrinsic beam features: the first-ring radius $k_\mathrm{L}$ sets the self-imaging period $z_s$, while the number of rings $N$ independently compresses the lateral width $w$ and depth-of-focus $\Delta z$. Implemented with a programmable SLM, the $\sqrt{n}$ radial discretization supports arbitrary complex overlays $\widetilde{A}(k_r,\chi)$, preserving topology (e.g., OAM) and enabling self-imaging of various spatial modes, including families of Gaussian modes and an Airy beam. Compared with the cylindrical Talbot analogue, the Montgomery carpet concentrates energy into tightly localized revivals ($\sim10-20$~\unit{\um} spots; $\Delta z\sim5~\mathrm{mm}$) across $z_s\approx30-100~\mathrm{mm}$, offering a compact, lensless route to dynamic multi-plane excitation with a reduced off-target dose. This work highlights the versatility of aperiodic self-imaging phenomena, whose distinctive axial intensity profiles may overcome limitations in conventional self-imaging -- such as optical tweezers, where extended intensity distributions of the Talbot effect pose challenges, and multi-plane fluorescence microscopy and two-photon, where minimizing off-target excitation and photodamage is critical. The Montgomery self-imaging technique may thus find applications across a broad range of fields, including microscopy, quantum computing, and quantum telecommunications. 


\textbf{Funding:} Work at Harvard was supported by the ONR MURI program, under award N00014-20-1-2450, and by the Air Force Office of Scientific Research (AFOSR) under award FA9550-22-1-0243. Work at UCF was supported by the US Office of Naval Research (ONR) under award N00014-17-1-2458, and the ONR MURI program under award N00014-20-1-2789.  L.A.H. was supported by the Los Alamos National Laboratory LDRD program grant 20251140PRD1. 

\bibliography{Montgomery}

\end{document}